\documentclass[preprint,showpacs,superscriptaddress,floatfix]{revtex4}
\usepackage{graphicx}
\usepackage{color}
\usepackage{amsmath}
\usepackage{amssymb}

\def\RR{\mathbb{R}}

\begin{document}

\title{Relevance of instantons in Burgers turbulence}

\author{Tobias \surname{Grafke}}
\affiliation{Department of Physics of Complex Systems, Weizmann
             Institute of Science, Rehovot 76100, Israel}
\author{Rainer \surname{Grauer}}
\affiliation{Theoretische Physik I, Ruhr-Universit\"at Bochum,
             Universit\"atsstr. 150, D44780 Bochum (Germany)}
\author{Tobias Sch\"afer}
\affiliation{Department of Mathematics, College of Staten Island, CUNY, USA}
\author{Eric \surname{Vanden-Eijnden}}
\affiliation{Courant Institute, New York University, 251 Mercer Street, New York, New York}
\date{\today}

\begin{abstract}
  Instanton calculations are performed in the context of stationary
  Burgers turbulence to estimate the tails of the probability density
  function (PDF) of velocity gradients. These results are then
  compared to those obtained from massive direct numerical simulations
  (DNS) of the randomly forced Burgers equation. The instanton
  predictions are shown to agree with the DNS in a wide range of
  regimes, including those that are far from the limiting cases
  previously considered in the literature. These results settle the
  controversy of the relevance of the instanton approach for the
  prediction of the velocity gradient PDF tail exponents. They also
  demonstrate the usefulness of the instanton formalism in Burgers
  turbulence, and suggest that this approach may be applicable in
  other contexts, such as 2D and 3D turbulence in compressible and
  incompressible flows.
\end{abstract}

\pacs{47.27.Ak, 47.27.E-, 47.27.ef, 05.40.-a}

\maketitle

The stochastically driven Burgers equation reads~\cite{burgers:1974}
\begin{equation} 
  \label{eq:burgers}
  u_t+uu_x-\nu u_{xx} = \eta,
\end{equation}
where $\eta$ is a white-noise forcing satisfying
\begin{equation}
  \label{eq:chi}
  \langle\eta(x,t)\eta(x',t')\rangle = \delta(t-t')\chi (x-x'),
\end{equation}
in which the spatial correlation $\chi(x)$ has characteristic length
$L$ and amplitude $\chi(0)=\chi_0$.  Besides having a wide range of
applications e.g. in the context of structure formation in the early
universe~\cite{arnold-zeldovich-etal:1982,shandarin-zeldovich:1989},
traffic flow \cite{chowdhury-santen-etal:2000}, growth processes
\cite{kardar-parisi-etal:1986}, etc. (see e.g. \cite{bec-khanin:2007}
for an overview), this equation has also gained considerable interest
as a toy-model to benchmark techniques for analyzing turbulence. This
is due mainly to the phenomenological simplicity of the solutions
to~\eqref{eq:burgers}: In stationary Burgers turbulence, velocity
perturbations with negative gradient evolve into shocks, while
positive gradients are smoothed out. The shocks have a dramatic
influence on the statistics of the velocity field: for example, they
are responsible for the anomalous scaling of the velocity increments
and they make the probability density function (PDF) of the velocity
gradient highly non-Gaussian. These features are signatures of
intermittency, the understanding of which has been the main issue in
turbulence theory~\cite{frisch:1995}.

In the context of Burgers turbulence, both the scaling of the right
tail of the velocity gradient PDF \cite{polyakov:1995,
  gurarie-migdal:1996, e-khanin-etal:1997, gotoh-kraichnan:1998} and
that of its left tail in the inviscid case \cite{e-vandeneijnden:2000,
  gotoh-kraichnan:1998, boldyrev-linde-polyakov:2004} are known. In
contrast, the scaling of the left tail in the viscid case
\cite{gotoh:1999, balkovsky-falkovich-etal:1997} remains more
controversial: In particular there is an inconsistency between
measurements of the exponent of the tail decay in direct numerical
simulations (DNS) \cite{gotoh:1999} and the predictions made
in~\cite{balkovsky-falkovich-etal:1997}. The latter were obtained
through approximations within the framework of the instanton method
\cite{gurarie-migdal:1996,falkovich-kolokolov-etal:1996,
  balkovsky-falkovich-etal:1997}, which is a field-theoretic approach
that has been used in hydrodynamic turbulence. Due to its
non-perturbative nature the instanton method is in principle
well-suited to study the probability and evolution of rare and extreme
events (i.e. the most singular/dissipative structures of the flow)
that are responsible for intermittency. This possibilty was also
confirmed recently by the numerical computation
\cite{chernykh-stepanov:2001,
  grafke-grauer-schaefer-vandeneijnden:2014} and successful
observation of instantons in actual Burgers turbulence
\cite{grafke-grauer-schaefer:2013}. The main aim of the present paper
is to investigate further the range of applicability of the instanton
method in this setup. In particular, we revisit the results of
\cite{balkovsky-falkovich-etal:1997}, where approximations were made
that permit to solve the instanton equations asymptotically and
predict that the left tail of the PDF is captured by a compressed
exponential with a given exponent. We show that these results are
backed up by numerical solutions to the instanton equations, but only
apply in the very far tail. Away from this tail, the approximations
made in~\cite{balkovsky-falkovich-etal:1997} fail and numerical
solution of the exact instanton equations shows that the PDF is no
longer a compressed exponential. These predictions are in agreement
with measurements from DNS over a wide range of gradients. This
explains why the DNS results in~\cite{gotoh:1999}, which were believed
to contradict the instanton predictions, are in fact consistent with
this approach.

We begin by nondimensionalizing~\eqref{eq:burgers}. If we measure
length in units of~$L$, time in units of~$L^2/\nu$, and (consistently)
velocity in units of~$\nu/L$, \eqref{eq:burgers} becomes
\begin{equation}
  \label{eq:4}
  u_t + u u_x - u_{xx} = \sigma\eta
\end{equation}
where $\eta$ satisfies~\eqref{eq:chi} with $\chi(x)$ having now
characteristic length 1 and amplitude $\chi(0)=1$, and we defined
\begin{equation}
  \label{eq:5}
  \sigma = \chi^{1/2}_0 L^2 \nu^{-3/2}\,.
\end{equation}
Up to boundary effects whose impact can be made negligible by making
the system size bigger, this parameter is the only control parameter
left in the system. It can be related to the Reynolds number
$\text{Re} = UL/\nu$ as $\sigma = \text{Re}^{3/2}$ if we use as
characteristic velocity $U = (\chi_0L)^{1/3}$ -- this velocity is the
root-mean-square velocity in the turbulent regime when the dissipation
scale $L_d = \nu^{3/4}\chi_0^{-1/4}$ is much smaller than $L$
(i.e. $\text{Re}$ and $\sigma$ are much bigger than 1) and an inertial
range develops.

%

The instanton method that we will use to analyze~\eqref{eq:4} relies
on Martin-Siggia-Rose/Janssen/de~Dominicis formalism
\cite{martin-siggia-rose:1973, dedominicis:1976, janssen:1976}. The
saddle point or instanton approximation can be made rigorous within
the framework of large deviation theory~\cite{varadhan:2008}. Within
this formalism the expectation of any observable $\mathcal{O}[u]$ of
the velocity field $u$ is represented as the path integral
\begin{equation}
  \label{eq:pathint}
  \langle \mathcal{O}[u] \rangle \propto \int Du\, \int D(ip)
  \mathcal{O}[u] \exp (-\sigma^{-2} I[u,p])
\end{equation}
where $\langle\cdot\rangle$ denotes expectation with respect to the invariant
measure of~\eqref{eq:burgers} and
\begin{equation}
  \label{eq:action}
  I[u,p] = \int_{-\infty}^0\left(\langle p, \dot u + u u_x - u_{xx}
  \rangle - \tfrac12 \langle p, \chi\star p \rangle\right) dt
\end{equation}
is the action functional for~\eqref{eq:burgers}. Here,
$\langle \cdot, \cdot \rangle$ is the $L_2(\RR)$-scalar product and
$\star$ denotes convolution. Since we are interested in strong velocity
gradients, we set
\begin{equation}
  \label{eq:obs}
  \mathcal{O}[u] = \exp(\sigma^{-2}\lambda u_x(0,0))
\end{equation}
Indeed, if
$S_*(\lambda) = \sigma^2 \ln \langle e^{\sigma^{-2} \lambda
  u_x(0,0)}\rangle$
and $p(a)$ denotes the PDF of $u_x(0,0)=a$, for large
$\sigma^{-2}|\lambda|$ we have
\begin{equation}
  \label{eq:1}
  S_*(\lambda) \equiv \sigma^2 \ln \int_{\RR} e^{\sigma^{-2} \lambda a} p(a) da \sim \max_{a}
  (\lambda a - S(a))
\end{equation}
where~$S(a) = - \sigma^2 \ln p(a)$ and we used Laplace's method to
estimate the integral.  This means that $S_*(\lambda)$ is the
Fenchel-Legendre transform of $S(a)$, which also implies that
\begin{equation}
  \label{eq:2}
  S(a) \sim \max_{\lambda}(\lambda a - S_*(\lambda)) 
\end{equation}
For large $\sigma^{-2}|\lambda|$ we can also relate $S_*(\lambda)$ and
$S(a)$ to the saddle point of the path integral in~\eqref{eq:pathint}
for the observable in~\eqref{eq:obs}, i.e. to the minimizer of the
action $I[u,p] - \lambda u_x(0,0)$. Specifically
\begin{equation}
  \label{eq:3}
  I[u^*,p^*] = \lambda  u_x^*(0,0)- S_*(\lambda) \sim S(u_x^*(0,0)) 
\end{equation}
where $(u^*,p^*)$ denote the minimizer, termed the \textit{instanton},
i.e. the solution to
\begin{equation}
  \label{eq:instanton}
  \begin{aligned}
    u_t + u u_x - u_{xx} &= \chi\star p\\
    p_t + u p_x + p_{xx} &= 0
  \end{aligned}
\end{equation}
with boundary conditions
\begin{equation}
  \label{eq:bndcond}
  u(t=-\infty) = 0,\qquad p(t=0) = -\lambda\delta'(x).
\end{equation}
The final condition for $p$ arises from incorporating the term
$- \lambda u_x(0,0)$ in the variational problem. By varying~$\lambda$,
we can access different values of the gradient, $u_x^*(0,0)$ and then
use~\eqref{eq:3} to compute $S (u_x^*(0,0)) $. Carrying on this
program therefore allows us to estimate $S(a)$ for different values of
$a$. Note that this function is independent of the control parameter
$\sigma$ (and hence the Reynolds number $\text{Re}=\sigma^{3/2}$)
since we have scaled this parameter out the instanton
equations~\eqref{eq:instanton}. Note also that the solution
to~\eqref{eq:instanton} subject to the boundary
conditions~\eqref{eq:bndcond} is also the most likely way by which a
large velocity gradient can occur in the flow. In particular, for
large negative $\sigma^{-2}\lambda$, the instanton gives the evolution
and the final configuration of the prototypical extreme Burgers shock,
and should therefore be comparable to results of DNS of the stochastic
Burgers equation. We will check this claim below.

\begin{figure}[t]
  \includegraphics[width=0.7\linewidth]{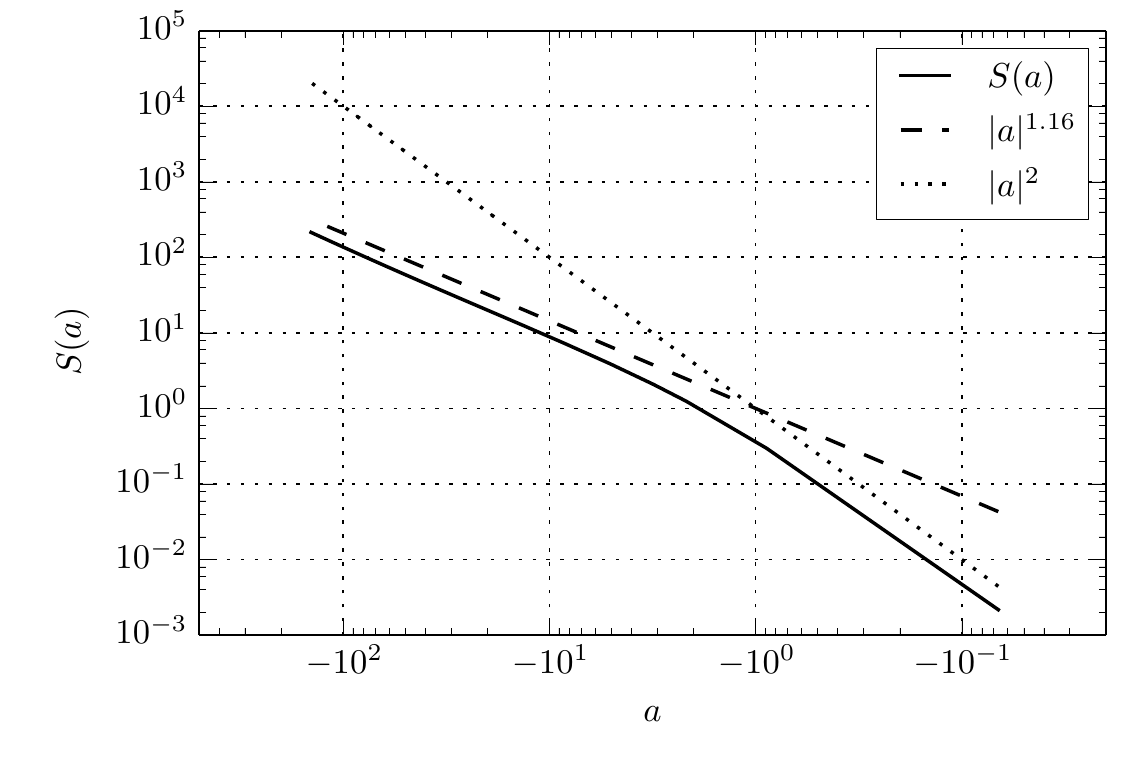}
  \caption{The parameter-free function $S(a)$ computed via solution of the
    instanton equations~\eqref{eq:instanton_geom} shown in log-log
    scaling for negative values of~$a$.}
  \label{fig:1}
\end{figure}
The instanton equations~\eqref{eq:instanton} were first integrated
numerically in~\cite{chernykh-stepanov:2001} and later on
in~\cite{grafke-grauer-schaefer:2013}.  These calculations turn out to be
challenging because the initial condition for~$u$
in~\eqref{eq:bndcond} is set at $t=-\infty$: in practice, we need to
set $t=-T$ for some large $T$, and check convergence by varying $T$,
but this requires taking larger and larger values of $T$ as
$|\lambda|$ increases. To overcome this problem, here we use the
approach proposed in~\cite{grafke-grauer-schaefer-vandeneijnden:2014}
building on works in~\cite{e-ren-vandeneijnden:2004,%
  heymann-vandeneijnden:2008,heymann-vandeneijnden:2008c,zhou-ren-e:2008}
and solve a reparametrized version of~\eqref{eq:instanton} in which
the physical time $t\in(-\infty,0]$ is replaced by an artificial,
reparametrized time $s\in[-1,0]$ defined in such a way that
$\| u_s\|_{\chi} = \text{const}$, where
$\|v\|_{\chi} = \sqrt{\langle v, v\rangle_{\chi}}$ with
$\langle u, v \rangle_{\chi} = \langle u, \chi^{-1} v \rangle$. After
reparametrization, the instanton equations~\eqref{eq:instanton} become
\begin{equation}
  \label{eq:instanton_geom}
  \begin{aligned}
    r u_s + uu_x - u_{xx} &=  \chi\star p \\ 
    r p_s + u p_x + p_{xx} &= 0
  \end{aligned}
\end{equation}
where $r = \| \frac12 u_{xx} - uu_x \|_{\chi} / \|u_s\|_{\chi} $,
subject to
\begin{equation}
  \label{eq:bndcond_geom}
  u(s=-1) = 0,\qquad p(s=0) = -\lambda\delta'(x).
\end{equation}
\begin{figure*}[tb]
  \includegraphics[width=1.0\linewidth]{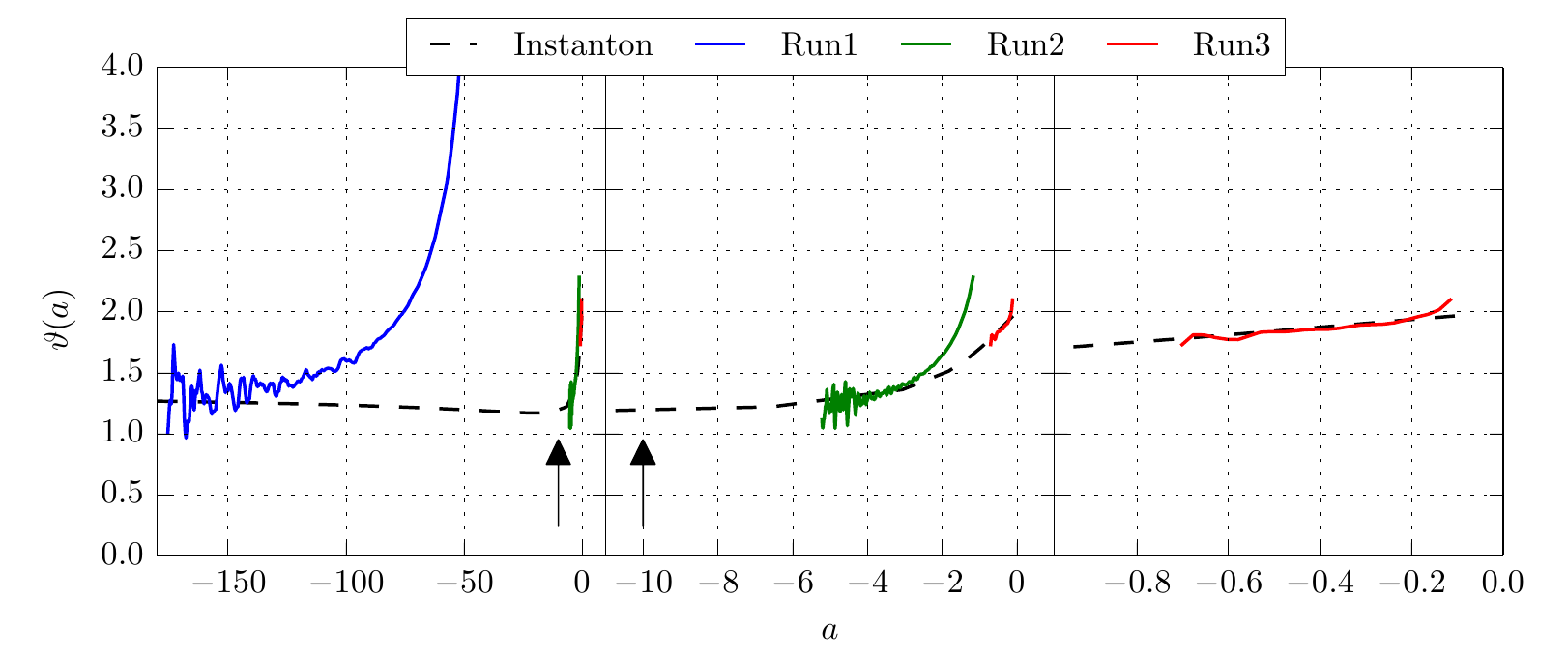}
  \caption{Local exponent $\vartheta(a)$ in~\eqref{eq:theta} obtained
    from the instanton calculation (dashed line) and compared the
    values estimated from DNS at the three different values
    of~$\sigma$ summarized in Table~\ref{table:runs}. }
  \label{fig:fas}
\end{figure*}

For the numerical simulations presented throughout this paper, we
chose
\begin{equation}
  \chi(x) = - \partial^2_{x} {\mathrm{e}}^{-x^2/2} = 
  (1-x^2){\mathrm{e}}^{-x^2/2}\,.
\end{equation}
The instanton equations were solved using a second order explicit
integrator in time with a time-step whose size is dictated by the
factor~$r$ in~\eqref{eq:instanton_geom} and can be approximated once
for several computations. We used fast Fourier transforms for all
spatial derivatives. This scheme was implemented as a GPU/CPU hybrid
code for speeding up the computations. The details of its
implementation, especially in terms of computational efficiency and
reduction of memory requirements, is discussed in
\cite{grafke-grauer-schindel:2014}. Here we simply note that, because
of the mixed initial and final boundary conditions
\eqref{eq:bndcond_geom}, algorithms computing transition probabilities
\cite{bouchet-laurie-zaboronski:2011, e-ren-vandeneijnden:2004,
  heymann-vandeneijnden:2008} with a known initial and final state are
not directly applicable in this setup.

\begin{table}[h]
\begin{tabular}{p{0.2\columnwidth}|p{0.2\columnwidth}|p{0.2\columnwidth}|c}
	 & \centering $\sigma$ & \centering $l_d$  & \#$T_L$ \\ \hline
\centering Run1 & \centering 17.21     & \centering 0.285  & $7.139 \cdot 10^5$ \\ 
\centering Run2 & \centering 1.70     & \centering 0.909  & $9.505 \cdot 10^8$ \\ 
\centering Run3 & \centering 0.52     & \centering 1.875  & $4.266 \cdot 10^6$    
\end{tabular}
\caption{Parameters of the  DNS at different values of~$\sigma$:
  driving amplitude, $l_d=L_d/L = \sigma^{-1/2}$: dissipation length, \#$T_L$: total
  number of integral times.
  \label{table:runs}}
\end{table}

The function $S(a)$ obtained by this method confirms the scaling
$S(a) \asymp a^3$ for large positive values of $a$. Here we focus on
large negative values of $a$, where the situation is more complex:
$S(a)$ is plotted against $a$ for $a<0$ in Fig.~\ref{fig:1} using a
log-log scaling. At first glance, it seems like $S(a) \asymp |a|^{2}$
in the core, then switches to $S(a) \asymp |a|^{\vartheta}$ with
$\vartheta\approx 1.16$ for larger negative values of $a$. This
exponent is not consistent with the theoretical prediction
$\vartheta _\infty=3/2$ obtained
in~\cite{balkovsky-falkovich-etal:1997}.  It is, however, very close
to the value $\vartheta=1.15$ that was measured by
Gotoh~\cite{gotoh:1999} in DNS $\textrm{Re}=2$ (i.e.
$\sigma \approx 2.83$). To explain the origin of this discrepancy, let
us take a closer at the function $S(a)$ and define the local exponent
\begin{equation}
  \label{eq:theta}
  \vartheta(a) := \frac{d\ln S}{d\ln |a|} = \frac{a}{S}\frac{dS}{da}
\end{equation}
It can be seen in Fig.~\ref{fig:fas} that this exponent keeps varying
slowly as $a$ decreases to larger negative values, indicating that
$S(a)$ is not yet a power-law for the values of $a$ plotted in
Fig.~\ref{fig:1}.  In fact, further calculations (not shown) indicates
that $\vartheta(a)\to 3/2$ as $a\to-\infty$, consistent with the
analytical prediction in~\cite{balkovsky-falkovich-etal:1997}. This,
however, happens for much larger values of $|a|$. At the same time,
the arrow in figure \ref{fig:fas} denotes the largest negative
gradient observed in Gotoh's DNS simulation Run1, demonstrating that,
in fact, his measured value is in remarkable agreement with the
instanton prediction for the local exponent at this value of the
gradient. Therefore, the results from \cite[Run1]{gotoh:1999} are
compatible with the statistics being dominated by instanton-like
events, albeit far from the limiting case where the approximation made
in~\cite{balkovsky-falkovich-etal:1997} apply.

This can be further confirmed by comparing these results to our own
massive DNS of~\eqref{eq:4}. These simulations were conducted with a
total of $2.6\times 10^{11}$ computational steps, amounting to about
$10^9$ large eddy turnover times in total, for various values of
$\sigma$, as summarized in Table~\ref{table:runs}. The local
exponent~$\vartheta(a)$ measured in these experiments is also shown in
Fig.~\ref{fig:fas} for three values of $\sigma$. As can been seen, in
each cases, the exponent estimated from the DNS eventually approaches
the instanton prediction, albeit at values that are not $3/2$. Note
the huge range of different gradients captured by this figure.

To assess the range of validity of the instanton predictions, it is
useful to plot the gradient PDF estimated from the DNS against
$Ce^{-\sigma^{-2} S(a)}$, where the constant $C$ is the normalization
factor of the PDF which is lost in the steepest descent
calculation. To make this comparison, we also renormalized the driving
amplitude~$\sigma$ for the more turbulent Run1 and Run2 to account for
the fluctuations that are not captured in the instanton calculation:
specifically $\sigma_1^\text{eff} = 0.83 \sigma_1$ and
$\sigma_2^\text{eff} = 0.92 \sigma_2$.  The results of these
calculations are reported in Fig.~\ref{fig:fas}, where we scaled
$u_x = a $ by its standard deviation $\sigma$ to make the ranges of
the different plots comparable. These graphs show how the tail of the
gradient PDF fattens as $\sigma $ increases.  They also show that the
instanton prediction always matches the estimate from the DNS if $|a|$
is large enough, and the smaller $\sigma$, the larger the range in
which agreement is observed. This is consistent with the fact that the
estimate in~\eqref{eq:1} relies on $\sigma^{-2} |\lambda|$ being
large, which becomes a more stringent requirement as $\sigma$ (and
hence the Reynolds number) increases. At the same time, since the left
tail of the gradient PDF fattens as $\sigma$ increases, the instanton
prediction does remain relevant to explain intermittency. In fact, if we
set $S(a) \sim C |a|^{\vartheta}$ in~\eqref{eq:1}, and assume that
$\vartheta$ is roughly constant, we see that the maximum is attained
at
\begin{equation}
  \label{eq:6}
  a_* = (\lambda/(C\vartheta))^{1/(\vartheta-1)} \gg 
  (\sigma^2/(C\vartheta))^{1/(\vartheta-1)}
\end{equation}
where the inequality indicates the range of gradients where the
instanton calculation will apply. Since it follows from~\eqref{eq:4}
that the standard deviation of $u_x=a$ is $\sigma$, this means that,
for large $\sigma$, the instanton method will capture gradients whose
amplitude is $\sigma^{(3-\vartheta)/(\vartheta-1)}$ times larger than
their standard deviation.

\begin{figure}[tb]
  \includegraphics[width=0.7\linewidth]{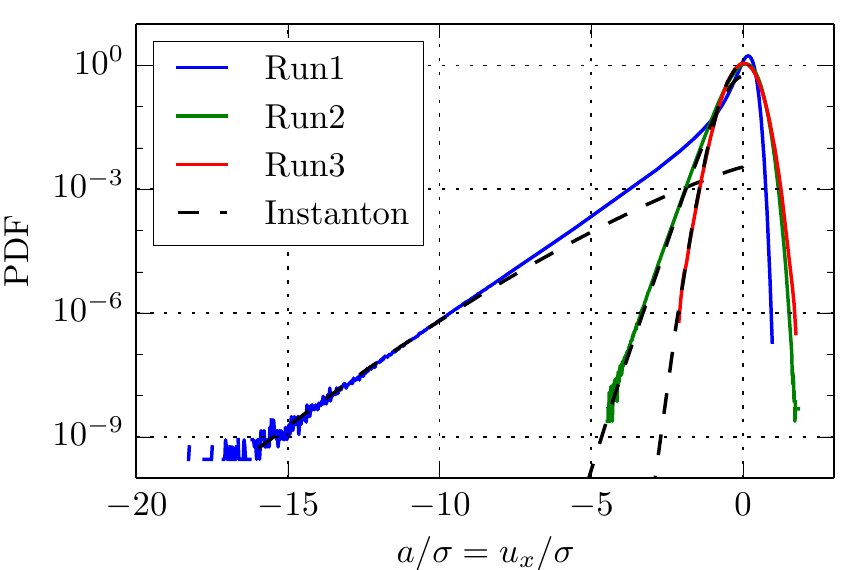}
  \caption{The gradient PDF obtained by the instanton method via
    $Ce^{-\sigma^{-2} S(a)}$ (dashed lines) is compared to the PDF
    estimated from DNS (solid lines) at the three different values
    of~$\sigma$ summarized in Table~\ref{table:runs}.  }
  \label{fig:allPDFs}
\end{figure}

In conclusion, the instanton approximation is able to reliably predict
scaling exponents of the velocity gradient PDF for rare events over a
broad range of values. Since the applicability of the method is
directly related to the Reynolds number Re, there is little hope of
measuring the limiting case of $\vartheta_\infty=\frac32$ in
DNS. Nevertheless, for moderate Re flows the tail scaling can be
estimated from the instanton, and for low Re the whole PDF can be
derived from the instanton configuration. This also answers the open
question raised in~\cite{gotoh:1999} of the applicability of the
instanton approach, and his measured exponent of $\vartheta=1.15$
agrees with our prediction of $\vartheta=1.16$ quite remarkably. The
major task for subsequent investigations is to include fluctuations
into the presented computations, which would permit predictions for
flows with higher Re, and to scale up these calculations to turbulent
flows in higher dimensions. Both aims seem achievable.

{\it Acknowledgments:} We thank Gregory Falkovich for helpful
discussions. The work of T.G. was partially supported through the
grants ISF-7101800401 and Minerva – Coop 7114170101. The work of R.G.
benefited from partial support through DFG-FOR1048, project B2. The
work of T.S. was partially supported by the NSF grant DMS-1108780. The
work of E.V.-E. was partially supported by NSF Grant No. DMS07-08140
and ONR Grant No. N00014- 11-1-0345.

\end{document}